\documentclass[aps,
twocolumn,
showpacs,amssymb,groupedaddress]{revtex4}

\usepackage{graphics,epsfig}
\usepackage{amsmath}
\usepackage{amsfonts}

\newcommand{\no}{\nonumber\\}
\newcommand{\be}{\begin{equation}}
\newcommand{\ee}{\end{equation}}
\newcommand{\ba}{\begin{eqnarray}}
\newcommand{\ea}{\end{eqnarray}}

\newcommand{\la}[1]{\label{#1}}

\def\tr{{\rm tr}}

\begin{document}
\title{Abnormal dilepton yield  from local parity breaking in
heavy-ion collisions}
\author{A.A. Andrianov$^{1,2}$\footnote{Corresponding author. E-mail: andrianov@icc.ub.edu}}
\author{ V.A. Andrianov$^{1}$}
\author{D. Espriu$^{2,3}$}
\author{X. Planells$^{2}$}
\affiliation{$^1$\ \  V.A.Fock Department of Theoretical Physics,  Saint-Petersburg State University, 198504, St.Petersburg, Russia}
\affiliation{$^2$\ \ Departament d'Estructura i Constituents
de la Mat\`eria and Institut de Ci\`encies del Cosmos (ICCUB) ,
Universitat de Barcelona, Mart\'\i \ i Franqu\`es 1, 08028 Barcelona, Spain}
\affiliation{$^3$\ \  CERN, 1211 Geneva, Switzerland}

\begin{abstract} We propose a novel explanation for the dilepton excess observed in heavy ion collisions at invariant
masses below 1 GeV. We argue that the presence of local parity breaking due to a time-dependent
isosinglet and/or isotriplet  pseudoscalar condensate
substantially modifies the dispersion relation of photons and vector mesons propagating in such a medium,
changing the $\rho$ spectral function and resulting in a potentially large
excess of $e^+e^-$ and $\mu^+ \mu^-$ with respect to the theoretical predictions based in a `cocktail' of known 
hadronic processes. Possible signatures to prove or disprove this effect are discussed.
\end{abstract}

\maketitle

During the last decade several experiments in heavy ion collisions  have indicated an abnormal yield of lepton pairs with
invariant mass $M < 1$ GeV in the region of small rapidities and moderate
transversal momenta \cite{ceres,phenix} (reviewed in \cite{lapidus,tserruya}).
This effect is visible only for collisions that are central or semi-central, and is present for $e^+e^-$ and also for
$\mu^+ \mu^-$. From a comparison to $p p$ and $p$-nucleus collisions it has been
established beyond doubt that such an
enhancement is a nuclear medium effect \cite{lapidus}. For the energies accessible at GSI (of few GeV per nucleon
in HADES experiments \cite{lapidus}) the effect was partially interpreted as being due to enhanced $\eta$ meson production
in proton-neutron scattering \cite{lapid1}. For the higher energies accessible at CERN SPS and
BNL RHIC (with $\sqrt{s_{NN}}$ ranging from 20 GeV to 200 GeV per nucleon in c.m., in CERES, HELIOS/3,
NA60 \cite{ceres} and PHENIX \cite{phenix} experiments)
the abnormal dilepton yield has not been yet explained
satisfactorily by known processes in hadronic physics\cite{lapidus,tserruya}.

Following \cite{phenix} we divide the range of dilepton invariant masses into high ( $M> 3.2$ GeV), low ( $M< 1.2$ GeV)
and intermediate. The low mass region (LMR) is in turn divided into LMR I with $M < 0.3$ GeV
and LMR II with   $0.3$ GeV $< M < 1.2$ GeV.
In LMR I the enhancement could possibly be explained by modifications of meson properties in nuclear
medium \cite{brown,rapp,cassing,zahed} as well as by proton-neutron scattering \cite{lapid1}. But in the LMR II
the $\rho$ meson, directly via $\pi\pi$ fusion or indirectly
through Dalitz processes, largely dominates and the in-medium effects of a dropping mass and/or broadening resonance 
seem unable to explain the spectacular dilepton enhancement by a 4 to 7 factor, depending
on $p_T$ and centrality (see \cite{heesrapp,renkr} for recent attempts).

In this letter we propose a radically different explanation of this enhancement. We suggest that the
effect may be a manifestation of
local parity breaking (LPB) in colliding nuclei due to the generation of a pseudoscalar, isosinglet or
neutral isotriplet, condensate whose magnitude depends on the dynamics of the collision. It has been suggested
that such a background could be due to the topological charge fluctuations leading to the so-called
Chiral Magnetic Effect (CME)\cite{kharzeev} studied by lattice QCD simulations \cite{lattice} and seemingly
detected in the STAR and PHENIX experiments at RHIC\cite{star}, although 
the issue is far from being settled. It might be also
related to pseudoscalar domain walls\cite{gorsky}. However the fact that the observed
dilepton excess is almost absent for peripheral collisions (where the CME should be more visible)
and maximized in cental collisions makes us believe that it may be due to the ephemeral
formation of a bona-fide thermodynamic phase where parity
is broken, a possibility that has been argued for in \cite{anesp}.

It has been shown in \cite{axion} that a pseudoscalar field slowly evolving in time drastically changes the
electromagnetic properties of the vacuum. In particular an energetic photon propagating in this background may decay on shell
into dileptons. This same mechanism extended to vector mesons is proposed here as the source
for the abnormal dilepton yield in the LMR, i.e. in the range $ M < 1.2$ GeV,
for centrality $0 \div 20\%$ and for moderate $p_T < 1$ GeV \cite{phenix}. In this letter we will concentrate in the LMR II,
and more specifically in the region around the $\rho$ and $\omega$ resonant contribution.

We shall assume that a time dependent but approximately spatially homogeneous background
of a pseudoscalar field $a(t)$ is induced at the
densities reached in heavy ion collisions and we will define a 4-vector related,
$\zeta_\mu  \simeq \partial_\mu a$, for later use. $a(t)$ could be either isosinglet or isotriplet
or even a mixture of the two, but detailed calculations will be presented for the case
of isosinglet background only.

The appropriate framework to describe
electromagnetic interactions of hadrons at low energies is the Vector Dominance Model (VDM)\cite{rapp,vmd}
containing the lightest vector mesons $\rho_0$ and $\omega$ in the $SU(2)$ flavor sector. We do not include $\phi$ meson,
as its typical mean free path $\sim 40$ fm makes it insensitive
to medium effects. Quark-meson interactions are described by
\ba
&&{\cal L}_{int} = \bar q \gamma_\mu V^\mu q;\quad  V_\mu \equiv - e A_\mu Q  +
\frac12 g_\omega  \omega_\mu \mathbf{I} + g_\rho \rho_\mu  \frac{\tau_3}{2},\no && (V_{\mu,a})
\equiv \left(A_\mu,\, \omega_\mu, \, \rho_\mu\equiv (\rho_0)_\mu\right), \label{veclagr}
\ea
where $Q= \frac{\tau_3}{2} + \frac16 \mathbf{I}$, $g_\omega \simeq  g_\rho \equiv g \simeq 6 $. These values
are extracted from vector meson decays. The Maxwell and mass terms are
\ba
&&\!\!\!{\cal L}_{kin} = - \frac14 \left(F_{\mu\nu}F^{\mu\nu}+ \omega_{\mu\nu}\omega^{\mu\nu}+
 \rho_{\mu\nu}\rho^{\mu\nu}\right)\\ &&\!\!\!{\cal L}_{mass} =m^2_V \mbox{\rm tr}( V_{\mu}V^\mu) = \frac12  V_{\mu,a}  m^2_{ab} V^\mu_b, \no  &&\!\!\! m^2_{ab} =
m_V^2\left(\begin{array}{ccccc}
\frac{10 e^2}{9g^2} & &-\frac{e}{3g} && -\frac{e}{g} \\
 -\frac{e}{3g}&& 1 && 0 \\
 -\frac{e}{g} && 0 && 1 \\
\end{array}\right),\ \mbox{\rm det}\left( m^2\right) = 0,\nonumber \label{vdm}
\ea
where  $m_V^2 = m^2_\rho = 2 g^2_\rho f_\pi^2\simeq m^2_\omega$ . This matrix reflects the VMD relations
at the quark level \cite{vmd,rapp}. Finally, in a pseudoscalar time-dependent background
the Lagrangian contains a parity-odd Chern-Simons (CS) term
\ba
&&{\cal L}_{CS}(k)\,= - \frac14 \varepsilon^{\,\mu\nu\rho\sigma}\, \tr{ \,\hat\zeta_\mu \, V_\nu(x)\, V_{\,\rho\sigma}(x)}\no
&&= \frac12 \tr{\,\hat\zeta \,\epsilon_{jkl}\, V_{j} \,\partial_k V_{l} }
= \frac12 \,\zeta \,\epsilon_{jkl}\, V_{j,a} \,N_{ab}\,\partial_k V_{l,b},
\ea
which additionally mixes photons and vector mesons due to LPB. For isosinglet pseudoscalar background
$e^2 \hat\zeta=\frac95\zeta\mathbf{I}$, and
the mixing matrix  reads
\ba N_{ab}\, \simeq\,  \left(\begin{array}{ccccc}
1 &  &-\frac{3g}{10e}& &-\frac{9g}{10e}\\
-\frac{3g}{10e}& & \frac{9g^2}{10e^2} & & 0 \\
-\frac{9g}{10e}& & 0 & &\frac{9g^2}{10e^2} \\
\end{array}
\right),\ \ \mbox{\rm det}\left( N\right) = 0 . \la{tab1}\ea
Remarkably, $ N \sim   m^2$. Simple order-of-magnitude considerations indicate that $\zeta\sim \alpha
\tau^{-1} \sim 1 $ MeV, taking  the time of formation of pseudoscalar condensate $\tau= 1$ fm and the value of 
condensate of order of $f_\pi$.

For isotriplet pseudoscalar background
$e^2 \hat\zeta=3\zeta\tau_3$, and the
corresponding CS matrix takes the form
\ba N^\pi_{ab} \,\simeq  \, \left(\begin{array}{ccccc}
1 & & -\frac{3g}{2e} & & -\frac{g}{2e} \\
-\frac{3g}{2e} & & 0 &  &\frac{3g^2}{2e^2}\\
-\frac{g}{2e}& & \frac{3g^2}{2e^2} & & 0 \\
\end{array}
\right),\quad \mbox{\rm det}\left( N^\pi\right) = 0 . \la{tab2}
\ea
The VMD coefficients in \eqref{tab1},\eqref{tab2} are obtained from the anomalous Wess-Zumino action \cite{truhlik}
and related to the phenomenology of radiative decays of vector mesons \cite{radec}.
The ratios of matrix elements for isotriplet condensate in \eqref{tab2} are in direct agreement with the experimental
decay constants for the processes
$\pi_0 \rightarrow \gamma\gamma,\quad  \omega \rightarrow \pi_0\gamma,\quad \rho_0 \rightarrow \pi_0\gamma$ \cite{truhlik}
and for the decay $\omega \rightarrow \pi\pi\pi$ \cite{weise} taken from \cite{pdg}. Likewise the elements in \eqref{tab1}
can be, in principle, estimated from the decays
$\eta \rightarrow \gamma\gamma,\quad \eta' \rightarrow \gamma\gamma,\quad\omega \rightarrow \eta\gamma,\quad \rho_0 \rightarrow \eta\gamma $.
However, phenomenologically there exists a strong $\eta_8 - \eta_0$ mixing effect 
finally resolved in the SU(3) flavor
scheme \cite{mixing}. Only the ratio of the decay widths $\omega \rightarrow \eta\gamma,\quad \rho_0 \rightarrow \eta\gamma $
is little sensitive to the mixing and confirms the off-diagonal elements of \eqref{tab1}.
In this letter we ignore the above mixing and restrict ourselves to SU(2).

The mass-shell equations for vector mesons read
\ba
&&K^{\mu\nu}_{ab} V_{\nu, b} = 0;\quad k^\nu\,V_{\nu, b} = 0, \no
&& K^{\mu\nu} \equiv g^{\mu\nu} (k^2 \mathbf{I} -  m^2) - k^\mu k^\nu \mathbf{I} -
i \varepsilon^{\,\mu\nu\rho\sigma}\,\zeta_\rho  k_\sigma  N ,
\ea
selecting out three physical polarizations for massive vector fields. In fact, these three polarizations contribute
into the vector field propagators as they couple to conserved fermion currents. The longitudinal
polarization $\varepsilon^\mu_L$ is orthogonal to $k_\mu$ and to the CS vector $\zeta_\mu$
\be
\varepsilon^\mu_L = \dfrac{\zeta^\mu k^2 - k^\mu (\zeta \cdot k)}{\sqrt{k^2 \big((\zeta \cdot k)^2 - \zeta^2 k^2\big)}},\quad
\varepsilon_{L}\cdot \varepsilon_L = - 1,
\ee
for $k^2 > 0$. The mass of this state remains undistorted.
The transversal (circular) polarizations $\varepsilon^\mu_\pm $ on the other hand satisfy
\ba
K^{\mu}_{\nu}\varepsilon^\nu_\pm = \Big(k^2 \mathbf{I} -  m^2 \pm \sqrt{(\zeta \cdot k)^2 - \zeta^2 k^2}\ N \Big) \varepsilon^\mu_\pm.
\ea

As previously mentioned we restrict ourselves to an isosinglet pseudoscalar background $a(t)$.
The spectrum can be found after the simultaneous diagonalization of matrices $ m^2, N $ and particularizing to the
case $\zeta_\mu \simeq (\zeta, 0,0,0) $
\ba
&&\!\!\!\! N\, = \, \mbox{\rm diag}\left[0,\,\frac{9g^2}{10 e^2},\, \frac{9g^2}{10 e^2}
+1\right] \sim \mbox{\rm diag} \left[0,\, 1,\, 1\right]\no
&& \!\!\!\! m^2\,  =\, m_V^2 \, \mbox{\rm diag} \left[0,\, 1,\,
1+ \frac{10 e^2}{9g^2}\right] \sim \mbox{\rm diag} \left[0,\, 1,\, 1\right] ,
\ea
namely
\be
k_0^2 - \vec k^2 = m_V^2 \pm \frac{9g^2}{10 e^2} \zeta |\vec k|\simeq m_V^2 \pm 360 \zeta |\vec k|\equiv m^2_{V,\pm} . \label{mvec}
\ee
Thus in the case of isosinglet pseudoscalar background the massless photons are not distorted when 
mixed with massive vector mesons.
In turn, massive vector mesons split into three polarizations with masses $m^2_{V,-} < m^2_{V,L}< m^2_{V,+}$.
This splitting unambiguously signifies local parity breaking as well as breaking of Lorentz invariance
due to the time-dependent background. For large enough $|\vec k| \geq 10 e^2 m^2_V / 9g^2 \zeta \simeq  m^2_V / 360 \zeta$
vector meson states with negative polarization become tachyons. However
their group velocity remains less than the light velocity \cite{ansol} provided
that $\zeta <  20 e^2 m_V / 9g^2 \simeq m_V / 180 \approx 4.3$ MeV.
For higher values of $\zeta$ the vacuum state becomes unstable, namely, polarization effects  give an imaginary part
for the vacuum energy.
Note that the position of resonance poles for $\pm$ polarized mesons is moving with wave vector $|\vec k|$ and
therefore they reveal themselves as "giant" resonances. The enlargement of the resonant region potentially
leads to a substantial enhancement of their contribution to dilepton production away from their nominal
vacuum resonance position.

The production rate of dileptons pairs mediated by $\rho$ mesons takes a form similar to the one 
given in \cite{rapp} but
with modified propagators due to LPB, according to our previous discussion
\begin{align}
\nonumber&\!\!\! \frac{dN}{d^4x dM}\simeq  c_\rho  \frac{\alpha^2 \Gamma_{\rho}m_{\rho}^2}{3 \pi^2 g^2 M^2}
\left (\frac{M^2-4 m_{\pi}^2}{m_{\rho}^2- 4 m_{\pi}^2}\right )^{3/2} \Theta(M^2 - 4 m_{\pi}^2)\\
&\!\!\!\times\sum_{\epsilon}\int_M^{\infty}dk_0\frac{\sqrt{k_0^2-M^2}}{e^{k_0/T}-1}
\dfrac{m_{\rho,\epsilon}^4 \left(1 +\frac{ \Gamma_\rho^2}{m_\rho^2}\right)}{\left (M^2-m_{\rho,\epsilon}^2\right )^2+
m_{\rho,\epsilon}^4\frac{ \Gamma_\rho^2}{m_\rho^2}}. \label{eleven}
\end{align}
Finite lepton mass corrections are not shown in (\ref{eleven}) but included in the fits. 
For $\omega$ mesons a similar expression is used but without the two pion threshold, characteristic of the 
dominant coupling of the $\rho$ to hadronic matter via two-pion fusion. A simple thermal average has been included
but we remark that the temperature $T$ is an effective one that may in fact depend on the range of $M$, $p_T$ and centrality. 
The coefficients $c_\rho, c_\omega $ parametrize in an effective way the total cross-sections for
vector meson creation. Because they are not known with precision in the present setting, particularly their off-shell values, 
the relative weights are used as free parameters in the hadronic `cocktail'\cite{ceres,na602,muonNA60}. This is the 
procedure used both by NA60 and PHENIX and we shall follow it here too. The usual `cocktail'
contains weights normalized to the peripheral collisions result (roughly agreeing with existing $pp$ and
$p$-nucleus data). For 
semi-central and central collisions, particularly at low $p_T$ the $\rho/\omega$ ratio needs to be enhanced
by a factor 1.6 in the case of NA60 \cite{ceres} or approximately 1.8 in PHENIX \cite{rykov}.  

A simple thermal average is appropriate only for central collisions  and moderate values of
$p_T$. No serious attempt will be made here to extrapolate to peripheral processes, but an appropriate
mixture of distorted and non-distorded vector mesons should be adequate. Likewise to 
include a proper description of the $p_T$ spectrum a more realistic statistical description would be needed.
Finally, for a proper comparison with the experimental results our expressions would have to be propagated 
through the experimental acceptances. Nevertheless the general features of the LPB-induced modifications
should be already visible in a simplified description.

NA60 has obtained accurate results for the $\rho$ spectral function\cite{ceres} by 
measuring the $\mu^+ \mu^-$ spectrum with unprecedent precision and by carefully
subtracting the contributions from the `cocktail' except the $\rho$ itself. The corresponding data
for central collisions is shown in Fig. 1 along with the contribution from the `cocktail' scaled by 
the 1.6 factor previously indicated and using $m_\rho=750$ MeV. The `cocktail' $\rho$ can
be obtained from (\ref{eleven}) by setting $\zeta=0$ 
and adjusting the effective temperature to $T\simeq 300$ MeV \cite{muonNA60}. However, even after optimization
of the constant to reproduce central $\rho$ production, the agreement 
of the data with the `cocktail' $\rho$ is poor,
reflecting the long standing problem of the insufficient dilepton yield. 
In the same figure we show, after adjusting the scale to reproduce the $\rho$ 
peak, the result for $\zeta\simeq 2$ MeV. 
Clearly the agreement is much better, particularly on the right of the
$\rho$ peak. On the left of the resonace there is a noticeable discrepancy for $M< 650$ MeV approximately; 
experimental points seem to be shifted upwards by a small constant amount with respect to
the prediction from LPB. It is revealing that
the upper kinematical limit for the Dalitz process $\omega\to \mu^+\mu^-\pi^0$ is 643 MeV. 

\begin{figure}
%[htp]
%\epsfxsize 8cm
\includegraphics
[scale=.24]{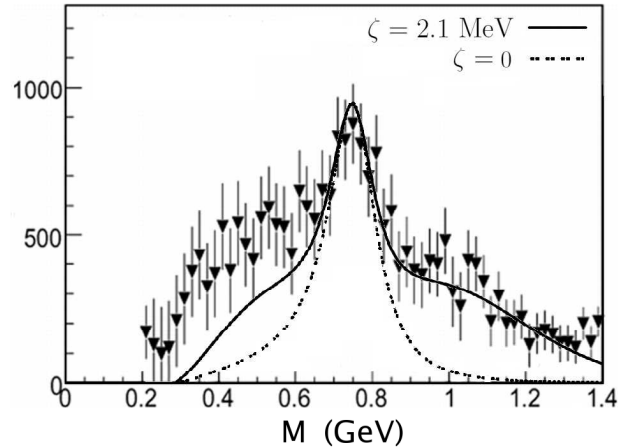}
\caption{The $\rho$ meson contribution into the dilepton production is shown for parity symmetric nuclear
matter $\zeta = 0$ and for local parity breaking with $\zeta = 2$ MeV compared to the NA60 measurement
for central collisons and all $p_T$. Both
contributions are normalized to produce the same on-shell cross-section for $\rho$ meson production, which
is fitted to the experiment. The effective temperature is 300 MeV.}\label{1}
\end{figure}

Important contributions to dilepton production at
lower invariant masses are the Dalitz processes $\pi_0 \rightarrow \gamma e^+e^-, \eta \rightarrow \gamma l^+l^- $,
nearly saturating the $l^+l^- $ production in the $M < 300$ MeV region, and $\omega\to l^+ l^- \pi^0$. The
latter has a partial lepton width nearly identical to the one of $\rho\to l^+ l^-$ and
thus expected to show a similar behavior. The Kroll-Wada
formula\cite{landsberg} includes the contribution of vector mesons to the previous Dalitz processes
and it remains valid in the
case of LPB provided that we replace the vector meson masses by the values in (\ref{mvec}) according to
the intermediate meson polarization ($L,\pm$). We have checked that this contribution
shows an enhancement but this along with all other hadronic processes relevant for dilepton production will be
discussed in a separate publication. Indeed NA60 has been able to itentify the individual contributions from
$\eta$ and $\omega$ Dalitz decays and found an enhancement, particularly for the latter\cite{na602}.

Note that the $\rho$ mean free path (1.3 fm) is much shorter than the expected
size of the hadronic gas fireball ($L=5 \div 10$ fm). The life-time of the fireball is comparable
$\tau_{FB} \simeq L$. The time spent by a ultra-relativistic resonance of width $\Gamma$ inside the fireball
is related to its mean free path $\sim 1/\Gamma$ in such a medium.
Therefore when $\Gamma \tau_{FB} \ll 1$  a tangible suppression
of the resonant enhancement $1/\Gamma^2$ arises. Around the distorted resonance peak a
crude estimation of this suppression gives $\sim \tau_{FB}^2\Gamma^2 $ in relative units.
This suppression affects all long lived particles such as the ones 
entering the Dalitz processes for the LMR I, and it is also 
relevant to $\omega$ meson decays away from the vacuum peak. It is also the ultimate reason why it is 
unnecessary to include the $\phi$ vector resonance in the discussion.
In our results, unless stated otherwise,
we have not included any modification of the $\omega$ propagator due to LPB.

Moving now to the PHENIX $e^+e^-$ data\cite{phenix,phenix2} we observe that it has a much poorer precision 
and a numerical subtraction of the `cocktail' to determine different individual contributions
is meaningless. For low values of $p_T$ and for central collisions the effective
temperature quoted by PHENIX lies in the region $100 \div 150$ MeV for the range of invariant masses
under consideration. Based on the standard `cocktail' we apply the normalization factor 1.8 already mentioned 
in order to approximately account for the correct $\rho/\omega$ ratio and use Eq. (\ref{eleven}) with
$\zeta \neq 0$ to include the effects of LPB. Fig. 2 shows the predicted $e^+ e^-$ yield for $z=1$ which 
is close to the optimal fit. The precision of the data does not allow for such clear cut conclusions
as in the case of NA60 but it is clear that LPB noticeably improves the agreement to the data as compared to the
`cocktail'. 

\begin{figure}
%[htp]
%\epsfxsize 8cm
\includegraphics
[scale=.24, trim=1.0cm 0 0 0 ]{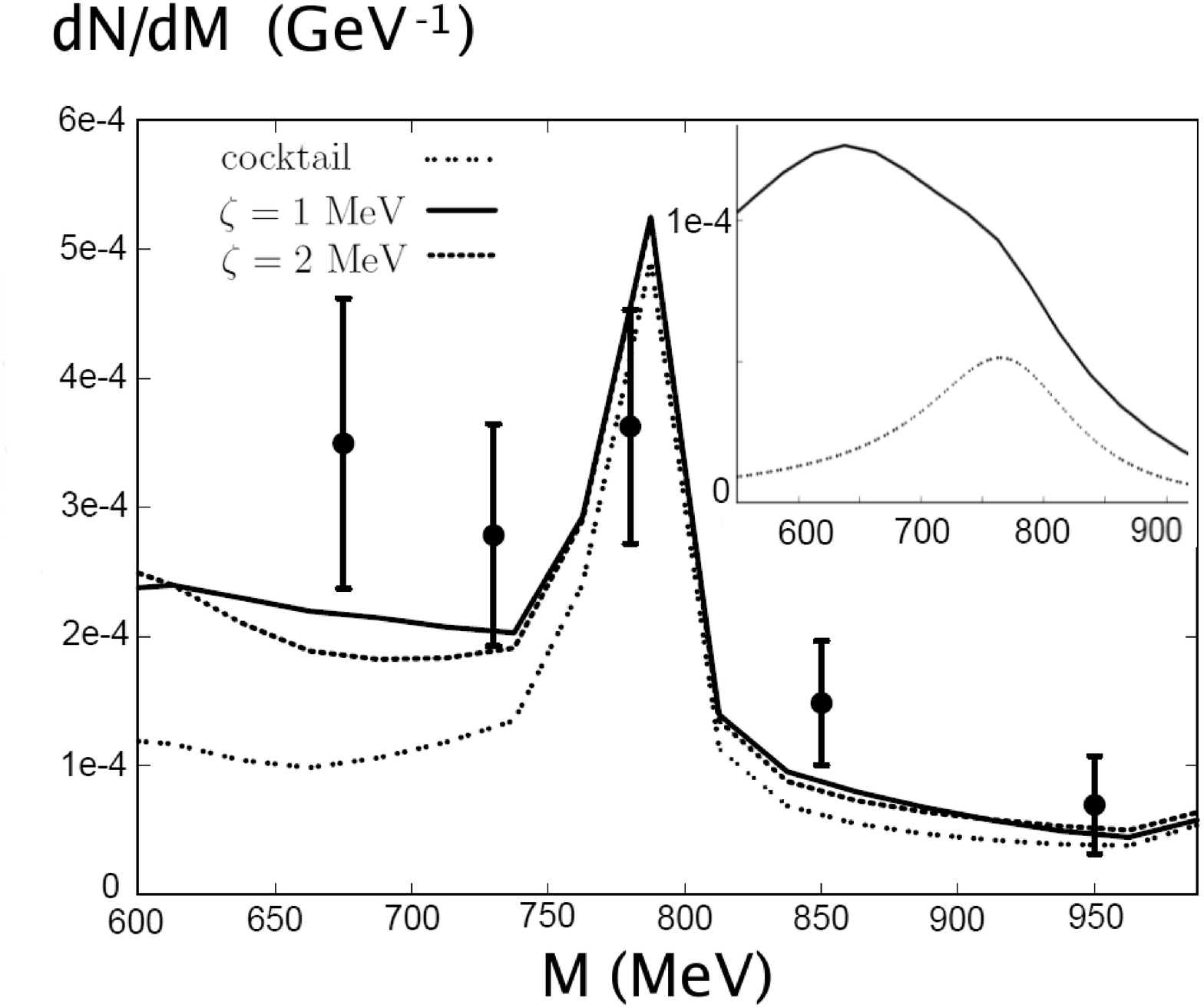}
\caption{The meson contribution to dilepton production is shown for parity symmetric nuclear
matter $\zeta = 0$ (discontinuous line) and for local parity breaking with $\zeta =1, 2$ MeV (solid line and
dotted line, respectively) compared to the PHENIX measurements for minimum bias, all $p_T$ events. 
The $\zeta=0$ line is just the `cocktail' contribution as quoted by PHENIX\cite{phenix2}.
The $\zeta\neq 0$ results from enhancing the $\rho/\omega$ ratio by a factor 1.8 without changing the $\omega$ normalization
w.r.t. the hadronic `cocktail'. The $\omega$ contribution itself includes a 20\% of LPB contribution ($\zeta=1$)
and 80\% pure `cocktail' ($\zeta=0$) to partially account for medium effects.
A reasonable fit is obtained
for $T=110$ MeV (compatible with PHENIX estimates for low $p_T$) and $\zeta=1$ MeV. Data for $M < 650$ MeV 
is not shown as the contribution from the Dalitz processes has not been included. Inset: the rho spectral
function for $\zeta=1$ MeV is compared to the cocktail one with the same assumptions.}\label{2}
\end{figure}

We would like to emphasize the extraordinary simplicity of the approach presented here. The fits 
presented use the values (effective temperatures, normalizations, etc. ) quoted by the experiment themselves.
The only free parameter is $\zeta$, which is expected to depend on the characteristics
of the collision. It should also be said clearly that the presence of LPB does not preclude 
other many body or in-medium corrections\cite{rapp}, as long as they do not represent double counting. 

We now summarize the signatures and  outline possible searches of local parity breaking.

Polarization: Dileptons produced for values of the invariant mass above and below the
$\rho+\omega$ pole are predominantly of opposite circular polarizations. Thus
one could search for a asymmetries among longitudinal
and transverse polarization for different $M$  in event-by-event measurements. These measurements may
reveal in an unambiguous way the existence of parity violation .

Distorted photons: At low energies massive vectors dominate the amplitudes but photons  
should also show a distortion induced by LPB. Those with `+' polarization exhibit different momentum
thresholds $\sim 4m^2_l / \zeta$ to show a resonant behavior for different dilepton species\cite{axion}.
Note that finite size suppresion is relevant for photons.

The nature of the condensate:  Mixing of photons with vector mesons is sensitive to isospin of pseudoscalar
condensate and therefore the fraction of distorted photon decays helps to disentangle its isospin content.

To summarize, in a time-dependent pseudoscalar background  massless photons of `+' polarization
and massive vector mesons behave as  giant resonances after averaging over thermal distribution.
For an isosinglet
pseudoscalar background in the framework of VMD only the massive vector mesons $\rho$ and $\omega$ propagators are distorted
due to mixing. We have computed their contribution and found that they naturally tend to
produce an overabundance of  dilepton pairs in the
$\rho+\omega$ resonance region. The modified $\rho$ spectral function shows features very
similar to the ones measured by NA60 in dimuon events. The PHENIX data is much better described
by this mechanism than by adjusting the standard hadronic `cocktail' in the $\rho+\omega$ resonance region.
At lower invariant masses
the Dalitz processes $\pi^0 \rightarrow \gamma e^+e^-, \eta \rightarrow \gamma l^+l^- $
and $\omega\to l^+ l^-\pi^0$ saturate
the hadronic contribution to $e^+e^-$ production and are enhanced
by the LPB-induced modifications on the vector meson propagators; detailed
results are deferred to a future publication.
The only free parameter is $\zeta$, characterizing the time variation of the pseudoscalar condensate.
A good fit to the data is obtained for natural values of $\zeta$. The possibility that the
LPB condensate is an isotriplet or an admixture of isotriplet and isosinglet has been discussed. 
Experimental signals of the manifestation
of LPB in heavy ion collisions have been suggested.

Thus local parity breaking seems capable of explaining in a natural
way the PHENIX/CERES/NA60 `anomaly' and searching for its manifestation in dilepton production represents 
an interesting challenge for some of the LHC collaborations.

We acknowledge the financial support from projects FPA2007-66665, 2009SGR502, CPAN (Consolider CSD2007-00042) and FLAVIANET.
A.\& V. Andrianov are supported also by Grants RFBR 09-02-00073-a and 10-02-00881-a and by SPbSU grant 11.0.64.2010.
We thank D. d'Enterria and J. Casalderrey for discussions and useful remarks.

\end{document}